\documentclass[twocolumn]{aastex631}
\usepackage{graphicx}
\usepackage{multirow}
\usepackage{txfonts}

\begin{document}

\title{Characterization of High-polarization Stars and Blazars with DIPOL-1 at Sierra Nevada Observatory}

\author[0000-0002-4241-5875]{J. Otero-Santos}
\altaffiliation{joteros@iaa.es}
\affiliation{Instituto de Astrof\'isica de Andalucía (CSIC), Glorieta de la Astronomía s/n, 18008 Granada, Spain}

\author[0000-0003-0186-206X]{V. Piirola}
\affiliation{Department of Physics and Astronomy, FI-20014 University of Turku, Finland}

\author[0000-0002-4131-655X]{J. Escudero Pedrosa}
\affiliation{Instituto de Astrof\'isica de Andalucía (CSIC), Glorieta de la Astronomía s/n, 18008 Granada, Spain}

\author[0000-0002-3777-6182]{I. Agudo}
\affiliation{Instituto de Astrof\'isica de Andalucía (CSIC), Glorieta de la Astronomía s/n, 18008 Granada, Spain}

\author[0000-0001-9400-0922]{D. Morcuende}
\affiliation{Instituto de Astrof\'isica de Andalucía (CSIC), Glorieta de la Astronomía s/n, 18008 Granada, Spain}

\author[0000-0002-9404-6952]{A. Sota}
\affiliation{Instituto de Astrof\'isica de Andalucía (CSIC), Glorieta de la Astronomía s/n, 18008 Granada, Spain}

\author[0000-0003-2036-8999]{V. Casanova}
\affiliation{Instituto de Astrof\'isica de Andalucía (CSIC), Glorieta de la Astronomía s/n, 18008 Granada, Spain}

\author[0000-0001-8074-4760]{F. J. Aceituno}
\affiliation{Instituto de Astrof\'isica de Andalucía (CSIC), Glorieta de la Astronomía s/n, 18008 Granada, Spain}

\author[0000-0002-1123-983X]{P. Santos-Sanz}
\affiliation{Instituto de Astrof\'isica de Andalucía (CSIC), Glorieta de la Astronomía s/n, 18008 Granada, Spain}



\begin{abstract}

We report here the performance and first results of the new multiband optical polarimeter DIPOL-1, installed at the Sierra Nevada Observatory 90 cm T90 telescope (SNO, Granada, Spain). DIPOL-1 is equipped with a plane parallel calcite plate and $\lambda$/2 retarder for modulating the intensity of two perpendicularly polarized beams, and a high readout speed CMOS camera that allows for fast, time-dense coverage. We characterize the performance of this instrument through a series of tests on zero- and high-polarization standard stars. The instrumental polarization in the Nasmyth focus was well determined, with a very stable contribution of 4.0806\% $\pm$ 0.0014\% in the optical $R$ band. For bright high-polarization standards ($m_{R}<8$) we reach precisions $<$0.02\% in polarization degree and 0.1\degr\ in polarization angle for exposures of 2$-$4~minutes. The polarization properties of these stars have been constrained, providing more recent results also about possible variability for future studies on some of the most used calibrators. Moreover, we have tested the capability of observing much fainter objects, in particular through blazar observations, where we reach a precision $<$0.5$-$0.6\% and $<$0.5\degr\ for faint targets ($m_{R}\sim16.5$) with exposures of $\sim$1 hour. For brighter targets ($m_{R}\sim14.5-15$), we can aim for time-dense observations with errors $<$$0.2-0.4$\% and $<$$1-1.5$\degr\ in 5-20~minutes. We have successfully performed a first campaign with DIPOL-1, detecting significant  polarized emission of several blazars, with special attention to the highest ever polarization degree measured from blazar 3C~345 at $\sim$32\%.
\end{abstract}



\section{Introduction}
The importance of polarization studies has experienced a significant growth during recent years. The increasing number of polarization observations and monitoring programmes \citep[e.g. RoboPol, see][]{ramaprakash2019} with the goal of studying all types of astrophysical objects introduces the need of developing more instruments that allow to study the broadband polarization\footnote{{A review of a large number of instruments and programmes} with polarization capabilities is presented by \cite{panapoulou2023}}. These needs have become even more urgent in the last two years after the launch of the Imaging X-ray Polarimetry Explorer (IXPE) satellite, {designed to measure high-energy polarization} \citep{weisskopf2016}. 

IXPE has been able to measure for the first time significant X-ray polarized emission from a large variety of astronomical objects: X-ray binaries \citep{podgorny2023}; pulsars \citep{suleimanov2023}; neutron stars \citep{dimarco2023}, active galactic nuclei \citep{ingram2023} and blazars \citep{liodakis2022}. However, to extract the full potential of these observations, simultaneous and complementary broadband polarimetric data are needed {\citep[e.g.][]{liodakis2022,ehlert2023,kim2024}}, covering all the electromagnetic spectrum. In such way, a complete interpretation of the physical mechanisms behind the polarized emission of astrophysical objects can be performed.

\begin{figure*}
    \centering
        \includegraphics[width=\columnwidth]{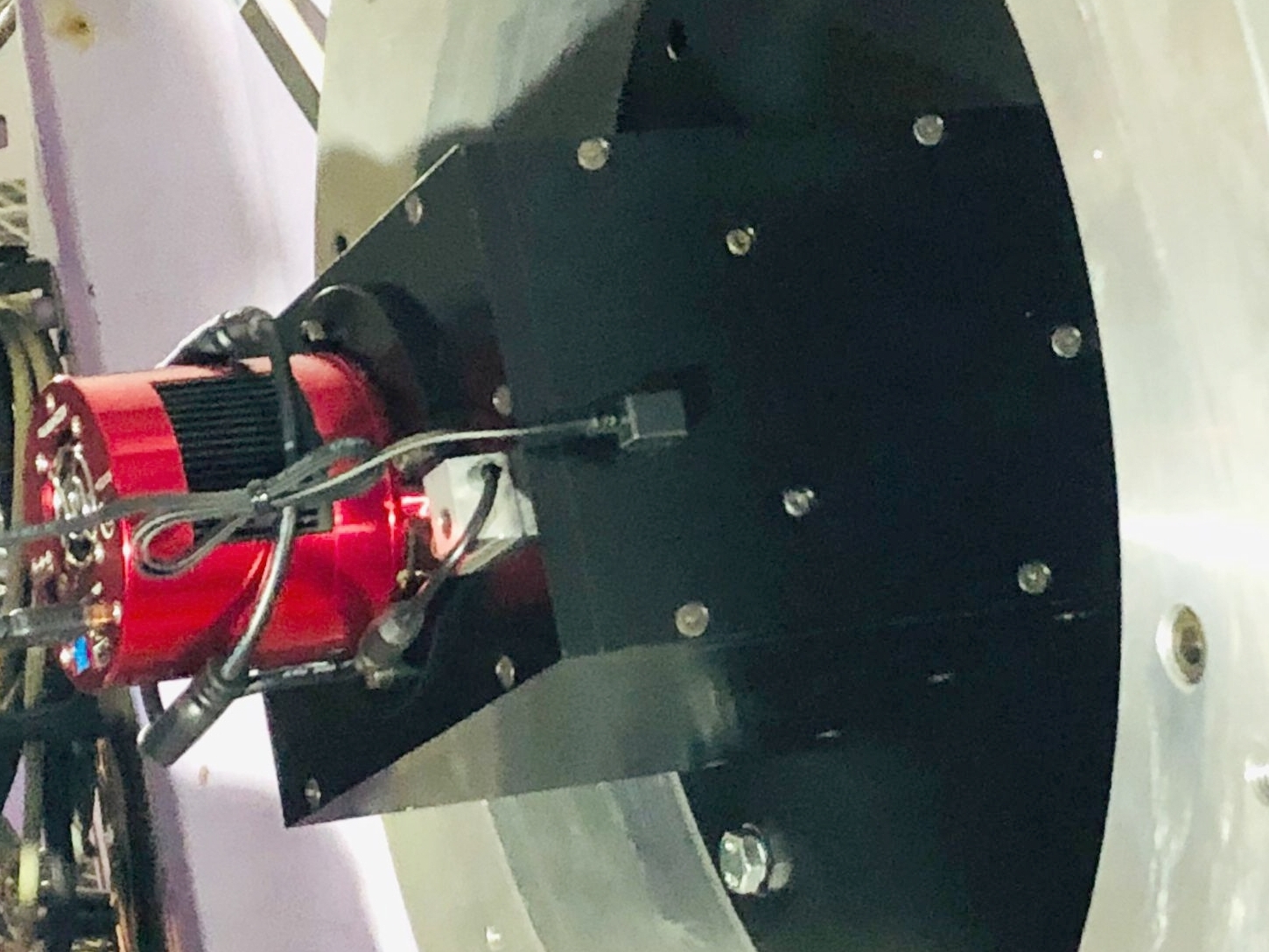}
        \includegraphics[width=\columnwidth]{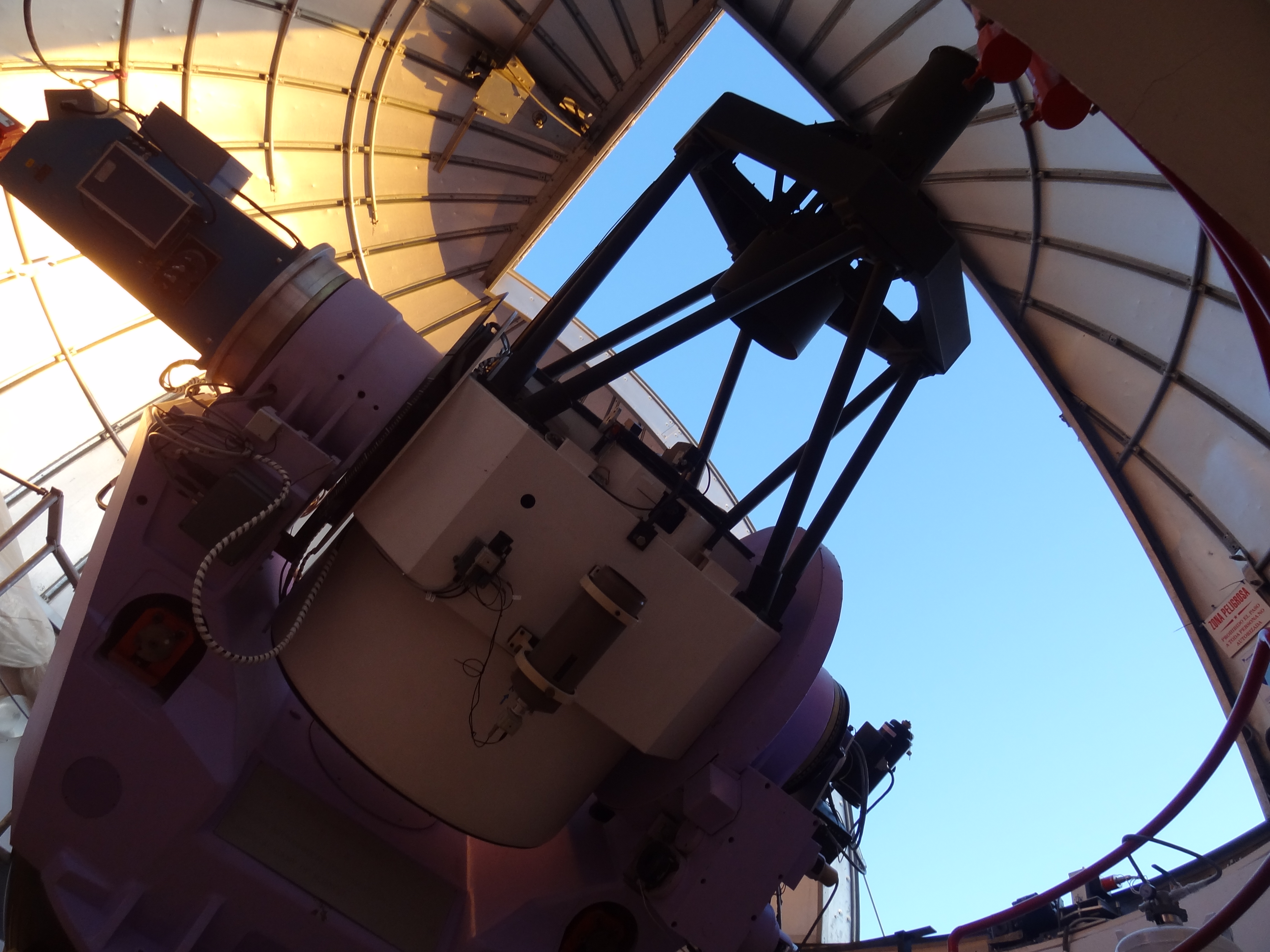}
    \caption{\textit{Left:} DIPOL-1 polarimeter with its CMOS camera on the T90. \textit{Right:} T90 telescope. \textit{Credit: IAA-CSIC.}}
    \label{fig:DIPOL1}
\end{figure*}

Under this premise, and with the aim of covering this need, we have installed the new optical polarimeter Double Image POLarimeter 1 (DIPOL-1), in September 2023 at the Sierra Nevada Observatory in Granada, Spain, operated by the Instituto de Astrof\'isica de Andaluc\'ia (IAA-CSIC). The instrument was designed, constructed and provided by the University of Turku (Finland) as part of its DIPOL programme\footnote{\url{https://sites.utu.fi/hea/research/optical-polarimetry/}}. Here we present the design and technical details of the polarimeter, as well as its instrumental calibrations and data reduction procedure. We also show some of the first results obtained with this instrument. The paper is structured as follows: in Section~\ref{sec2} we describe in detail the design and technical specifications of the polarimeter, including its optical system, camera and hardware, and the telescope used; in Section~\ref{sec3} we present the instrument control software; the data reduction procedure is detailed in Section~\ref{sec4};  the calibrations and performance tests are included in Section~\ref{sec5}; and some first results are shown in Section~\ref{sec6}. Closing remarks on the instrument and its performance are included in Section~\ref{sec7}.

\section{Design of the Polarimeter}\label{sec2}
The  DIPOL-1 instrument follows a similar --- yet more simplified --- design to other versions of DIPOL polarimeters such as DIPOL-2 and DIPOL-UF models \citep[see e.g.][]{piirola2014,piirola2020,piirola2021}. Figure~\ref{fig:DIPOL1} shows a photograph of DIPOL-1 installed on the T90. {The instrument has a field of view (FoV) of 9.25\arcmin~$\times$~6.30\arcmin\ and a pixel scale of 0.134~\arcsec/pixel.} 

\begin{figure}
        \includegraphics[width=\columnwidth]{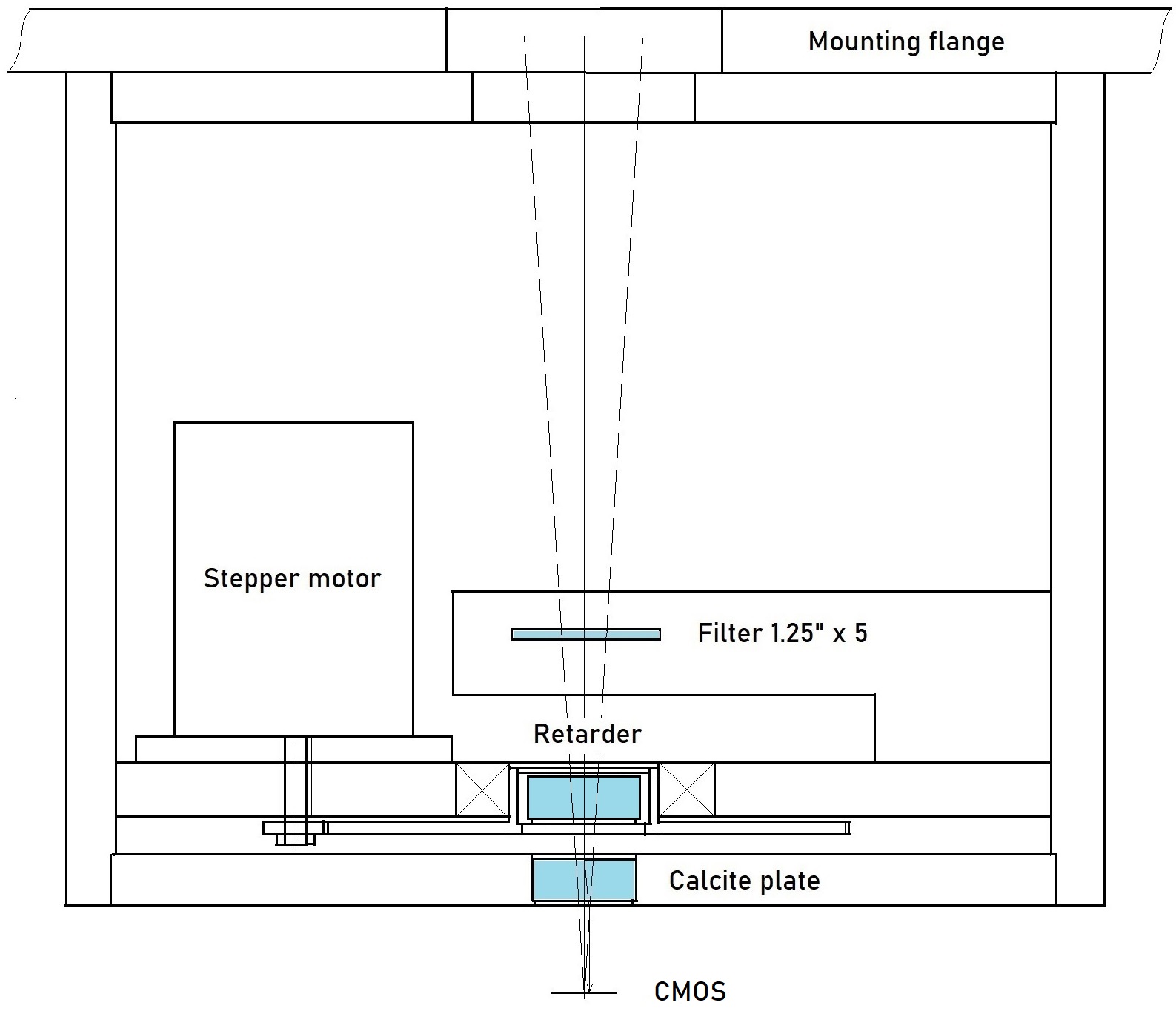}
    \caption{Scheme of the DIPOL-1 polarimeter. Rotatable superachromatic $\lambda$/2 retarder plate modulates the relative intensities of the two polarized beams produced by the calcite crystal, with an amount proportional to the degree of linear polarization of the incoming radiation. The fluxes of the two polarized stellar images are recorded with a highly sensitive cooled CMOS detector.}
    \label{fig:DIPOL-head}
\end{figure}

In Figure~\ref{fig:DIPOL-head} we represent a schematic diagram of the polarimeter head. The filter wheel accommodates five 1.25" diameter filters. The retarder, {a Super-Achromatic $\lambda$/2 plate APSAW-5\footnote{\url{http://astropribor.com/super-achromatic-quarter-and-half-waveplate/}} (400-700 nm) from Astropribor with thickness 7~mm}, is rotated at 22.5$^{\circ}$ intervals by a stepper motor, and modulates the intensities of the two perpendicularly polarized beams produced by the plane parallel calcite plate, with an amplitude proportional to the degree of polarization of the incoming radiation. At each position angle of the $\lambda$/2 plate an exposure is made of the double image of the target. The separation of the {ordinary (o-)} and {extraordinary (e-)} images is 1.0 mm, which corresponds to 28.65\arcsec in the focal plane of the T90 telescope. This is sufficient even in bad seeing conditions ($\gtrsim$2\arcsec) and also allows defocusing, which is useful for high S/N observations of bright stars to avoid saturation by spreading the light onto larger number of pixels. {All optical elements have an anti-reflecting MgF2 coating.}

{The oblique refraction introduces some astigmatism in the e-image. However, this image is round when the telescope is focused half-way between the astigmatic line foci. At this focal setting, the o-image is slightly out of focus, and the full-width half maximum (FWHM) values of the two images are equal. Nevertheless, we note that atmospheric seeing dominates over this issue, as the effect from the calcite plate is $<$0.5\arcsec - 1\arcsec, depending on the focal length of the telescope. Typical seeing values at the T90 telescope are $1.5-2$\arcsec.}

{Some polarimeters have noticeable position-dependent instrumental polarization, which arises from reflections in the strongly curved lenses of the re-imaging optics. There are no lenses at all in DIPOL-1, only plane-parallel optical elements. Therefore, it has the advantage of being free from position-dependent instrumental polarization.}

The high readout speed CMOS camera (ZWO ASI 294 MM Pro) minimizes the loss of time between exposures. Typically the individual exposure times range from 1 sec for bright calibration sources to 40 sec for the faintest targets. Observations are always made in cycles of 16 exposures, corresponding to a full rotation (360\degr) of the retarder. This minimizes systematic errors due to possible misalignment or non-parallelism of the rotating optical component and dust particles on the retarder surface. {The performance of the camera was found to be extremely good for the required precision, with no noticeable noise patterns such as ``popcorn noise''.}

High throughput sharp cutoff $RGB$ filters (Baader) with equivalent wavelengths of 650 nm, 535 nm, and 450 nm, and FWHM of 100 nm, 82 nm, and 114 nm, respectively, are currently installed. For calibrating the large ($\sim$4 \%) wavelength dependent instrumental polarization introduced by the 45\degr\ Nasmyth {aluminium coated} mirror, sharp cutoff filters are essential. The $\lambda$/2 plate is superachromatic in the wavelength range 400 -- 700 nm.

The T90\footnote{For more details on the T90 characteristics, see \url{https://www.osn.iaa.csic.es/page/telescopio-90-cm}} telescope (Figure~\ref{fig:DIPOL1}) has a Ritchey-Chrétien configuration and is equipped with two Nasmyth foci, one of which was used for the installation of DIPOL-1. A summary of its main characteristics is presented in Table~\ref{tab:t90}. It has a tracking with an accuracy $>$1$\arcsec$ on timescales of $\sim$1 minute. Moreover, for long exposures and large series of images, it is equipped with an off-axis autoguiding system that minimizes the drifting effect of the images towards the edges, allowing for a very precise tracking of the object.

\begin{deluxetable}{lcr}
\tablecaption{Technical characteristics of the T90 telescope.\label{tab:t90}}
\tablehead{
\multicolumn{3}{c}{T90 telescope characteristics} 
}
\startdata
Aperture & & 90 cm \\
Mount & & Equatorial \\
Configuration & & Ritchey-Chrétien \\
Focal ratio & & f/8 \\
Focal distance & & 7200 mm \\
Scale & & 35 $\mu$m/$\arcsec$\\
Field of View (no vignetting) & & 26.4$\arcmin$\\
\enddata
\end{deluxetable}

\section{Instrument Control}\label{sec3}

\subsection{Control Hardware}

The stepper motor for $\lambda/2$ plate rotation (Trinamic PD3-110-42), has an 
integrated control board communicating with the control computer 
(IPC-607 fanless mini-PC) via serial (RS-232) line. Reference switch is searched and 
the retarder driven automatically to the starting position always when the polarimetry 
script is started. Internal encoder provides means to check the actual position of the
$\lambda/2$ plate after each 22.5\degr\ rotation, before the next exposure is started.
Filter wheel is controlled via USB from the camera, incorporated into the MaxIm DL 
VBScript main control programme, which also sends the binary commands to the stepper 
motor and starts exposures for the desired number of measurement cycles.  

\subsection{Control Software}
The control software that allows all the DIPOL-1 elements to interact between each other 
is written in VBScript, applying MaxIm DL\footnote{\url{https://diffractionlimited.com/product/maxim-dl/}} 
scripting commands for the camera and filter wheel control and AxSerial COM port software 
package for communicating with the stepper motor by binary commands. The necessary parameters 
for the observation, target name and folder path, exposure time, and the filter, are specified in a simple text file given as requested by the script in a dialog box. Such scripts have already been used with success in other versions of DIPOL polarimeters \citep[e.g. DIPOL-2, see][]{piirola2014, piirola2020}.

\section{Data Processing and Reduction}\label{sec4}
DIPOL-1 observations require standard calibrations, i.e. bias and dark subtraction, and flat field correction \citep{berdyugin2019}. Bias and dark exposures are taken at the beginning of each observing night. Sky flats on the other hand can be taken on a sparser basis (e.g. once per week). {Flats are taken for one full rotation of the retarder plate (16 images) and the master flat is created through a median filtering procedure.} 

\begin{figure}
        \includegraphics[width=\columnwidth]{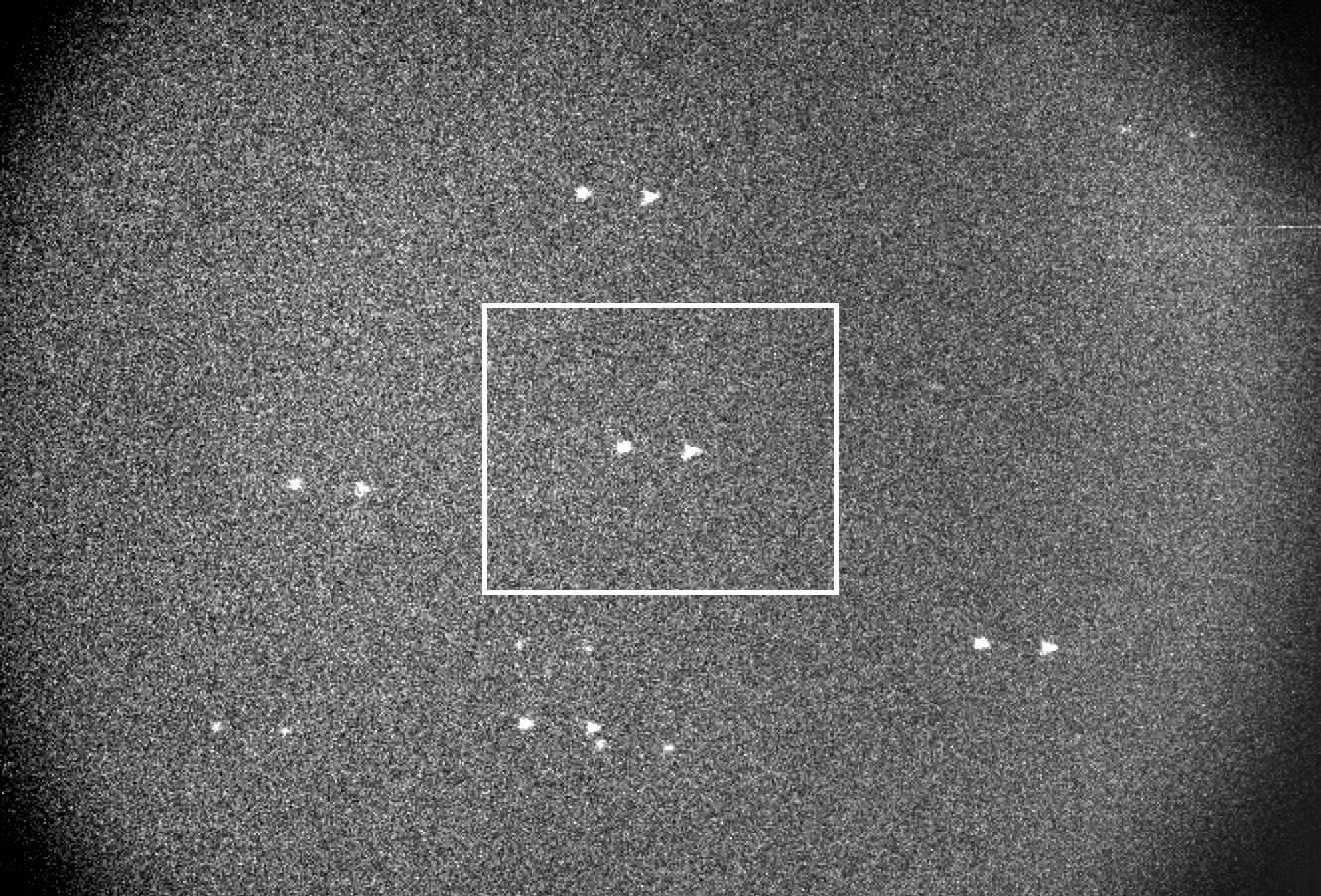}
    \caption{{Raw image of the blazar S4~0954+65 observed by DIPOL-1. The white rectangle corresponds to the sub-field saved and used in the polarimetry analysis. Slight amount of coma is due to telescope optics misalignment at the time of the observations. Flatfield features are enhanced by high contrast.}}
    \label{fig:raw_image}
\end{figure}

{Images taken for polarimetry correspond to a central image sub-frame. This is for optimizing the amount of data per night, due to the large amount of images saved and therefore, of storage capacity needed. An example of a raw image is shown in Figure~\ref{fig:raw_image}, where this central region is highlighted.} 

During the first stage of data reduction, the corresponding bias and dark images are averaged, calculating the mean bias and dark. These averaged images are used for performing the bias and dark subtraction to the object exposures. After that, the flat field correction can be applied. {From our extensive tests we have found that} the effect of the flat field correction has a minimal impact on the derived polarization parameters \citep[see also e.g.][]{witzel2011}. Therefore, this step can be skipped with no significant repercussion on the final results.

\begin{figure}
        \includegraphics[width=\columnwidth]{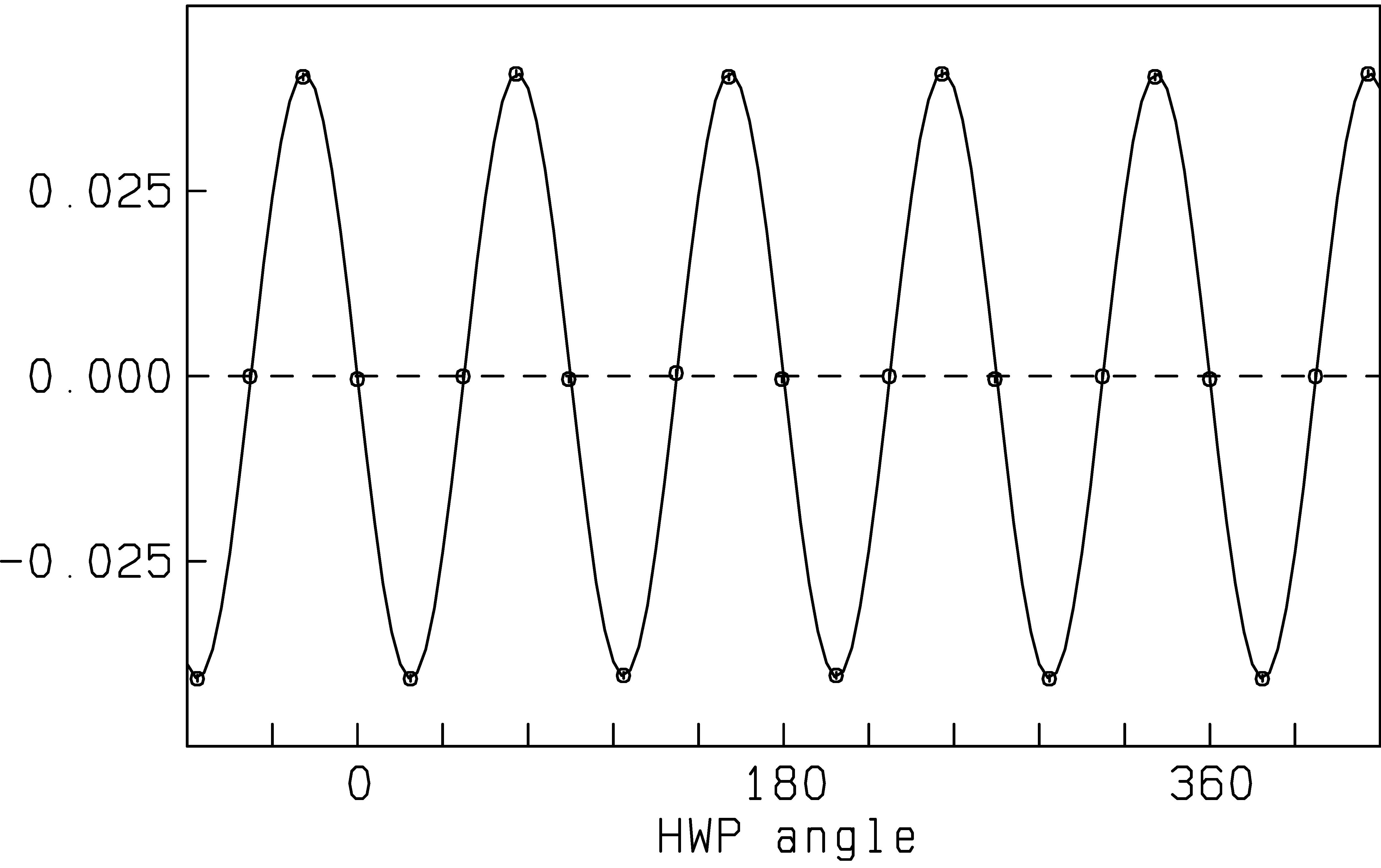}
    \caption{{Modulation curve example for the $\lambda$/2 plate rotation, averaged from 128 $\times$ 2 second exposures of the zero-polarization star HD~219623, in terms of the normalized flux differences $(F_o - F_e)/(F_o + F_e)$. Error bars are smaller than the plotting symbol. The curve shows Fourier fit with up to 4th harmonics included. The amplitude corresponds to the $\sim$4\% instrumental polarization produced by the Nasmyth mirror (M3). The power in the harmonics 1-3 is negligible (smaller by a factor $>$200 compared with that in the 4th). The difference in the transmission coefficients of the o- and e-beams is automatically calibrated out in the reduction procedure.}}
    \label{fig:aperture_photometry}
\end{figure}

After calibrations, the resulting images are analyzed to derive the polarization characteristics of the source. The brightness difference between the o- and e-rays is obtained through a differential aperture photometry method. This is done for each of the images at the $\lambda$/2 plate 16 positions, resulting in a sine-like variation of the relative brightness (see Figure~\ref{fig:aperture_photometry}). This step can be performed with the MaxIm~DL software. The Stokes parameters $Q$ and $U$ from each 4 successive exposures made at 22.5$^{\circ}$ intervals of the $\lambda$/2 plate are calculated from the intensity ratios \citep[see][]{berdyugin2019}. Then, all the individual Stokes parameters are averaged to derive the final Stokes parameters and thus, the polarization degree and angle values. This is done with a 2$\sigma$ weighted average algorithm, giving a lower importance to clear outliers and fluctuations introduced by, for instance, the effect of clouds or cosmic ray events in the camera \citep{kosenkov2017,piirola2020}. Moreover, points with a deviation $>$3$\sigma$ are given a zero weight and thus, rejected from the averaging procedure. 

The aperture radius used for performing the photometry of the o- and e-images, as well as the radius of the annulus used for the background estimation are optimized until reaching the result with the lowest statistical uncertainty. Finally, the instrumental polarization (see Section~\ref{sec5} for further details on its calculation) is removed, yielding the final values of the Stokes parameters $Q$ and $U$, as well as the polarization degree, $P$, and angle, $\theta$.

DIPOL-1 observations are carried out in such way that typically for each object, 8 observing cycles are performed, i.e. 8 series of 16 images, one per each $\lambda$/2 plate position (128 images in total). This provides significant flexibility in selecting the time intervals used for performing the weighted average and thus, to compute the Stokes parameters, the polarization degree and polarization angle. Therefore, one may iterate until reaching a balance between the precision of the results (in terms of the derived uncertainties) and time resolution. For bright targets, several bins may be computed, allowing for a denser coverage. On the other hand, faint targets may require averaging points on a nightly basis in order to obtain precise measurements of the polarization \citep[see for example][]{kosenkov2017,veledina2019,kosenkov2020}.

The reduction algorithm, already used in other DIPOL models \citep[for instance DIPOL-2, for more details see][]{piirola2020}, is implemented in a \textsc{Fortran} routine. The reduction procedure on \textsc{python} programming language as part of the automatic polarimetric reduction pipeline IOP4\footnote{\url{https://github.com/juanep97/iop4}} has also been implemented \citep{escudero2023}.

{DIPOL-1 images can be used also for performing photometry. However, since the images used for polarimetry correspond only to a central sub-frame of the FoV, in most cases there are no calibrators within this sub-field. Therefore, full frame images are taken for complementary photometry, although this not being the main purpose of the instrument.}

\section{Calibrations and Performance Tests}\label{sec5}
During the commissioning stage, we have performed observations of several zero-polarization stars from \cite{piirola2020} in order to determine the instrumental polarization. We have chosen a sample of nearby stars with very low interstellar polarization contribution and different spectral types (F-G-V), to avoid possible intrinsic polarization contributions as much as possible. The calculation of the instrumental polarization is particularly important in this case, as the instrument is mounted on a Nasmyth focus, that is expected to introduce a larger instrumental polarization than Cassegrain foci {\citep{anche2018}}, not easily available for this telescope. The estimations of the instrumental Stokes parameters and instrumental polarization, as well as the calibrations of the position angle zero point for the $RGB$ bands are reported in Table~\ref{tab:instrumental_polarization}. {These values are similar to those measured for instance in the Nasmyth focus of the Telescopio Nazionale Galileo \citep[TNG, see][]{giro2003}. However, a direct comparison is not straightforward, as different mirror coatings lead to different instrumental polarization and wavelength dependence \citep{wiersema2018}.} 

We note that during September 28-29, 2023, a cleaning of the T90 mirrors took place, leading to a slight change of the instrumental polarization. Therefore, we report this contribution for the pre-cleaning and post-cleaning periods. {We also stress that thanks to the T90 equatorial mount, this instrumental polarization is not a function of the parallactic angle, which makes reductions much easier than for alt-azimuth telescopes \citep[e.g.][]{giro2003,wiersema2018}.}

\begin{deluxetable}{lcccccc}
\tablecaption{Estimated instrumental polarization Stokes parameters, instrumental polarization degree, and polarization angle zero-point correction for different optical bands.\label{tab:instrumental_polarization}}
\tablehead{
  \colhead{Band} &
  \colhead{$Q_{i}$ (\%)} &
  \colhead{$U_{i}$ (\%)} &
  \colhead{m.e. (\%)} &
  \colhead{$P_{i}$ (\%)} &
  \colhead{$\theta$ (\degr)}
}
\startdata
\multicolumn{6}{c}{Pre-cleaning} \\
$R$ & 0.0578 & -3.7710 & 0.0016 & 3.7714 & 44.5 \\
$G$ & 0.3215 & -3.4448 & 0.0030 & 3.4600 & 42.3 \\
$B$ & -0.3603 & -3.8401 & 0.0046 & 3.8570 & 47.7 \\
\tableline
\multicolumn{6}{c}{Post-cleaning} \\
Band & $Q_{i}$ (\%) & $U_{i}$ (\%) & m.e. (\%) & $P_{i}$ (\%) & $\theta$ (\degr)\\ \tableline
$R$ & -0.0138 & -4.0806 & 0.0014 & 4.0806 & 45.1 \\
$G$ & 0.2572 & -3.7608 & 0.0026 & 3.7695 & 43.0 \\
$B$ & -0.5626 & -4.1434 & 0.0026 & 4.1814 & 48.9 \\
\enddata
\end{deluxetable}

We have also tested the stability of these estimations, with special dedication to the $R$ band owing to its much higher use, by measuring the polarization properties of the sample of zero-polarization stars after removing the instrumental polarization component. The results of the Stokes parameters for these stars after correcting this contribution are reported in Table~\ref{tab:zero_polarization_standards}. We observe that all values, except for the star HD 187013, are very close to a zero polarization measurement, typically $<$0.02\%, ensuring the stability of the instrumental polarization obtained from observing this star sample. The star HD 187013 clearly shows some intrinsic polarization, $p \sim0.06\%$, detected on two different nights, and was excluded from the instrumental polarization determination. For the post-cleaning period, we have tested the stability of the new instrumental polarization measurement with observations performed after MJD~60218. As seen in Table~\ref{tab:zero_polarization_standards}, these updated estimates have also found to be completely stable. In both cases we have conducted observations of zero-polarization standards at different zenith distances and positions on the sky, testing the stability of the instrumental contribution for different pointings of the telescope. This is clearly visible from the results of the non-polarized star HD~219623, observed for 8 consecutive nights at different zenith distances (between 17\degr\ and 58\degr), all observations resulting in nearly 0\% polarization, as expected from a good, stable calibration.

We note that the errors of 0.003-0.005\% for a single zero-polarization star observation 
($\sim$4 min total exposure) correspond to an uncertainty of  0.02\degr\ -- 0.04\degr\
in the position angle of the $P\sim4$\% instrumental polarization. The scatter from 
values obtained from different stars (and nights) is larger, with standard deviation 
$\sim$0.3\degr. This is partially due to small amount of interstellar and/or 
intrinsic polarization in the calibration stars, and to the uncertainty of the 
reference point determination for the initial position of the $\lambda/2$ plate. 
It is recommended to observe a null standard each time when the polarization 
script is started, and take the target measurements without exiting from the script 
meanwhile. Actually, this also serves as an accurate check of the zero-point correction
for position angles into the equatorial system, since the Nasmyth mirror 
in an equatorial telescope polarizes the light in the N-S direction ($\theta=0\degr$).

{The telescope mirrors M1 and M2 also contribute to the net instrumental 
polarization. However, we have found that typically in a Cassegrain 
telescope the instrumental polarization is in the level of 0.01\%. 
For a 0.05\% instrumental polarization of M1+M2 the effect in 
position angle of the total value would be $<$0.4\degr. In any case, 
standard star measurements are needed for the initial checks and 
long-term monitoring of the instrumental constants.}

\begin{deluxetable}{lccccccccccccc}
\tablecaption{Stokes parameters $Q$ and $U$ and $1\sigma$ errors m.e. in the $R$ band for a sample of zero-polarization standards \citep{piirola2020} after correcting for the effect of instrumental polarization, observed with DIPOL-1 at T90.\label{tab:zero_polarization_standards}}
\tablehead{
  \colhead{Star} &
  \colhead{$m_{V}$} &
  \colhead{\begin{tabular}[c]{@{}c@{}}Date\\ (MJD)\end{tabular}} &
  \colhead{$Q$ (\%)} &
  \colhead{$U$ (\%)} &
  \colhead{m.e. (\%)}
}
\startdata
HD 32923 & 5.00 & 60260.13$^{a}$ & 0.0007 & 0.0495 & 0.0068 \\
HD 152598 & 5.33 & 60211.79$^{b}$ & -0.0025 & 0.0176 & 0.0032 \\
HD 162917 & 5.76 & 60210.79$^{b}$ & 0.0035 & -0.0067 & 0.0047 \\
HD 173667 & 4.19 & 60207.82$^{b}$ & -0.0020 & 0.0027 & 0.0030 \\
 & & 60210.79$^{b}$ & -0.0036 & 0.0083 & 0.0028 \\
HD 184960 & 5.73 & 60207.33$^{b}$ & 0.0045 & -0.0078 & 0.0054 \\
 & & 60210.30$^{b}$ & 0.0034 & -0.0173 & 0.0042 \\
HD 185395 & 4.48 & 60210.31$^{b}$ & 0.0059 & -0.0088 & 0.0033 \\
HD 187013 & 4.99 & 60207.37$^{b}$ & -0.0546 & 0.0333 & 0.0052 \\
 & & 60210.32$^{b}$ & -0.0629 & -0.0236 & 0.0044 \\
HD 191195 & 5.85 & 60210.33$^{b}$ & -0.0012 & -0.0090 & 0.0039 \\
HD 193664 & 5.93 & 60242.91$^{a}$ & -0.0192 & -0.0024 & 0.0071 \\
HD 219623 & 5.45 & 60222.92$^{a}$ & -0.0008 & -0.0074 & 0.0029 \\
 & & 60223.05$^{a}$ & -0.0036 & 0.0019 & 0.0031 \\
 & & 60223.17$^{a}$ & 0.0044 & 0.0057 & 0.0033 \\
 & & 60223.84$^{a}$ & -0.0026 & 0.0112 & 0.0037 \\
 & & 60223.96$^{a}$ & 0.0044 & -0.0158 & 0.0033 \\
 & & 60224.12$^{a}$ & -0.0026 & 0.0071 & 0.0040 \\
 & & 60224.84$^{a}$ & -0.0024 & -0.0069 & 0.0034 \\
 & & 60224.96$^{a}$ & -0.0033 & -0.0027 & 0.0030 \\
 & & 60225.17$^{a}$ & 0.0053 & 0.0104 & 0.0031 \\
 & & 60225.83$^{a}$ & -0.0024 & 0.0039 & 0.0036 \\
 & & 60225.92$^{a}$ & 0.0034 & -0.0067 & 0.0031 \\
 & & 60226.16$^{a}$ & -0.0016 & 0.0469 & 0.0036 \\
 & & 60226.89$^{a}$ & -0.0066 & 0.0291 & 0.0035 \\
 & & 60227.02$^{a}$ & 0.0049 & -0.0127 & 0.0036 \\
 & & 60227.14$^{a}$ & 0.0017 & -0.0124 & 0.0038 \\
 & & 60227.98$^{a}$ & 0.0044 & 0.0028 & 0.0033 \\
 & & 60228.07$^{a}$ & 0.0001 & 0.0147 & 0.0034 \\
 & & 60228.13$^{a}$ & -0.0049 & -0.0175 & 0.0032 \\
 & & 60228.92$^{a}$ & 0.0038 & 0.0397 & 0.0036 \\
 & & 60229.03$^{a}$ & -0.0026 & -0.0286 & 0.0036 \\
 & & 60229.13$^{a}$ & -0.0014 & -0.0100 & 0.0046 \\
 & & 60229.92$^{a}$ & -0.0033 & 0.0000 & 0.0032 \\
 & & 60230.03$^{a}$ & -0.0018 & 0.0145 & 0.0032 \\
 & & 60230.16$^{a}$ & 0.0051 & -0.0227 & 0.0031 \\
\enddata
\tablecomments{$^{a}$Observations performed after the mirror cleaning. \\
$^{b}$Observations performed before the mirror cleaning.}
\end{deluxetable}

\begin{deluxetable*}{lcccccccccc}
\tablecaption{Sample of high-polarization standards observed with DIPOL-1 at T90.\label{tab:high_polarization_standards}}
\tablehead{
\multirow{3}{*}{Star} & \multirow{3}{*}{\begin{tabular}[c]{@{}c@{}}Date\\ (MJD)\end{tabular}} & \multicolumn{2}{c}{$R$} & & \multicolumn{2}{c}{$G$} & &\multicolumn{2}{c}{$B$} & \multirow{3}{*}{Refs.} \\  \cline{3-4}  \cline{6-7}  \cline{9-10}  
&   &     \colhead{$P$ (\%)}       &     \colhead{$\theta$ (\degr)}   &   &    \colhead{$P$ (\%)}       &     \colhead{$\theta$ (\degr)}   &   &      \colhead{$P$ (\%)}     &      \colhead{$\theta$ (\degr)}   &  
}
\startdata
   BD +59 389        &    60255.14$^{a}$       &    $6.468 \pm 0.018$       &    $100.1 \pm 0.1$   &    &     --      &    --       &    & -- &   --  &  1,2  \\   \hline
   HD 19820        &    60255.13$^{a}$       &    $4.648 \pm 0.021$       &    $114.2 \pm 0.1$   &    &     --      &    --       &    & -- &   --  &  1,2  \\   \hline
  HD 161056        &    60210.80$^{b}$       &    $3.946 \pm 0.014$       &    $68.3 \pm 0.1$   &    &     --      &    --       &    & -- &   --  &  1,2  \\   \hline
HD 204827       &     60207.88$^{b}$    &     $5.056 \pm 0.020$      &      $60.3 \pm 0.1$   &  &   $5.456 \pm 0.014$      &      $60.5 \pm 0.1$   &      &   $5.484 \pm 0.032$      &      $60.6 \pm 0.2$     &   1,2,3   \\  
    &   60210.83$^{b}$       &    $4.986 \pm 0.017$       &    $59.9 \pm 0.1$   &    &     --      &     --      &    &    --   &   --  &      \\ 
      &    60212.95$^{b}$     &   $5.021 \pm 0.019$      &      $60.2 \pm 0.1$   &    &     --      &     --      &    &    --   &   --  &      \\ 
    &    60213.13$^{b}$     &   $5.024 \pm 0.013$      &      $60.2 \pm 0.1$  &    &     --      &     --      &    &    --   &   --  &      \\ 
       &    60213.85$^{b}$     &   $4.983 \pm 0.017$       &    $59.8 \pm 0.1$   &    &     --      &     --      &    &    --   &   --  &      \\ 
      &    60214.09$^{b}$     &   $4.969 \pm 0.017$       &    $60.1 \pm 0.1$   &    &     --      &     --      &    &    --   &   --  &      \\  
      &    60242.82$^{a}$     &   $5.008 \pm 0.014$       &    $61.4 \pm 0.1$   &    &     --      &     --      &    &    --   &   --  &      \\ 
  &    60242.88$^{a,c}$     &   $5.001 \pm 0.044$       &    $61.2 \pm 0.3$   &    &     --      &     --      &    &    --   &   --  &      \\
      &    60243.05$^{a}$     &   $5.060 \pm 0.018$       &    $61.3 \pm 0.1$   &    &     --      &     --      &    &    --   &   --  &      \\\hline
\enddata
\tablecomments{{References:} (1) \cite{piirola2021}; (2) \cite{schmidt1992}; (3) \cite{turnshek1990}.\\ $^{a}$Observations performed after the mirror cleaning. \\ $^{b}$Observations performed before the mirror cleaning. \\ $^{c}$Data affected by high passing clouds.}
\end{deluxetable*}

A total of four high-polarization standard stars were observed during the commissioning stage to evaluate the accuracy of the estimated instrumental polarization contribution and zero point of the polarization angle $\theta$ of DIPOL-1 \citep{piirola2021}. For one of these four stars, HD~204827, several observations where performed before and after the mirror cleaning to check the stability of the measurements taken by the instrument. The results of the polarization degree and polarization angle derived for the observed high-polarization standards are shown in Table~\ref{tab:high_polarization_standards}. As proven by our results, the estimates derived for these high-polarization standards are well in agreement with previous values reported in the literature \citep{piirola2021}, and are completely stable over time, even after the mirror cleaning and instrumental polarization change.

Finally, as for the zero-polarization stars, we have tested the stability of these results for different pointings by observing these high-polarization standards at different zenith distances over the night. The perfect example of this stability is HD~204827, the most observed high-polarization standard so far. We have observed this star at zenith distances ranging from $\sim$20\degr\ and $\sim$60\degr. All the values derived at different zenith distances are well in agreement within $<$$\pm$0.05\% in polarization degree. Moreover, all results derived for the full sample of high polarization standards, chosen due to their distribution in the sky, are also compatible with those reported in the literature. Therefore, no effects of instrumental polarization or systematic errors are introduced in the analysis and further results due to the pointing of the telescope.


\section{First Results}\label{sec6}
DIPOL-1 was installed and commissioned at the T90 on 2023 September 19-22, and has been regularly taking data after its installation. With the observations of high-polarization standards we have tested the performance and stability of DIPOL-1 on targets of known, constant polarization degree and angle. All measurements are in fairly good agreement with previous estimates of the polarization standards {(within $\lesssim$0.05\%)}, reported in the different references indicated in Table~\ref{tab:high_polarization_standards} \citep{turnshek1990,schmidt1992,piirola2021}. The first DIPOL-1 observations also provide more updated measurements of these standards, that will be useful for future calibrations and polarimetric studies within the community.

We observe minor differences with regard to the most recent results from \cite{piirola2021} with the DIPOL-UF (Ultra Fast) polarimeter installed at the 2.5-m Nordic Optical Telescope (NOT) of $\sim$0.05-0.1\% in polarization degree and $\sim$0.5\degr\ in polarization angle in the $R$ band. Moreover, from the results presented in Table~\ref{tab:high_polarization_standards}, we observe that very high precisions of $\sim$0.01$-$0.02\% and $\sim$0.1\degr\ in polarization degree and angle are achieved for bright targets ($m_{V}<8$) in observing times $\sim$2$-$4~minutes. Most of the observations of high-polarization standards have been performed in the $R$ band owing to the much extensive use of this band by the ongoing polarimetry programmes at the T90. Nevertheless, we observe roughly the same precision {($\sim$0.1\%)} for the $G$ band with respect to the polarization values reported for HD~204827 by \cite{schmidt1992}. For the $B$ band we observe a slightly larger deviation from the values reported by \cite{schmidt1992, piirola2021} of 0.2\% polarization degree and $\sim$1.5\degr\ polarization angle. However, small variations of the polarization of this star have been claimed in the past \citep[e.g.][]{dolan1986}, which could explain the difference. Indeed, the values reported here are still within the possible variations presented by this reference high-polarization star according to \cite{bastien1988}. Therefore, all our results are well in agreement with previous measurements and completely stable.

\begin{deluxetable*}{lcccccccc}
\tablecaption{Polarization properties of a sample of blazars from the TOP-MAPCAT blazar list observed with DIPOL-1 at T90.\label{tab:blazars_polarization}}
\tablehead{
\multirow{3}{*}{Blazar} & \colhead{\multirow{3}{*}{\begin{tabular}[c]{@{}c@{}}R.A.\\ (hh:mm:ss)\end{tabular}}} & \colhead{\multirow{3}{*}{\begin{tabular}[c]{@{}c@{}}Dec.\\ (dd:mm:ss)\end{tabular}}} & \multirow{3}{*}{$m_{R}$$^{a}$} & \colhead{\multirow{3}{*}{\begin{tabular}[c]{@{}c@{}}Exp. time$^{b}$\\ (s)\end{tabular}}} & \colhead{\multirow{3}{*}{\begin{tabular}[c]{@{}c@{}}Date\\ (MJD)\end{tabular}}} &  \multicolumn{2}{c}{$R$} \\  \cline{7-8}    
&  & & &  & &   \colhead{$P$ (\%)} & \colhead{$\theta$ (\degr)} 
}
\startdata
J0211+1051     & 02:11:13.18 &   +10:51:34.80   &  14.5 & 20 &   60212.10       &    $6.301 \pm 0.152$       &    $166.1 \pm 0.7$      \\ 
    & &     &   & 30 &   60213.97       &    $6.060 \pm 0.154$       &    $2.5 \pm 0.7$      \\ \hline
3C 66A & 02:22:39.61 & +43:02:07.80 &  15 & 30 &     60211.12    &     $12.459 \pm 0.192$      &      $176.4 \pm 0.4$      \\  
 & & &   & 20 &     60212.05    &     $12.319 \pm 0.187$      &      $177.5 \pm 0.4$      \\  
 & & &   & 30 &     60213.01    &     $11.655 \pm 0.155$      &      $178.2 \pm 0.4$      \\
 & & &   & 30 &     60214.02    &     $9.600 \pm 0.177$      &      $177.1 \pm 0.5$     \\   \hline
3C 84          & 03:19:48.16 & +41:30:42.11  & 12  & 10 &   60212.13       &    $1.214 \pm 0.114$       &    $74.6 \pm 2.7$      \\ 
         &  &  &   & 20 &   60213.06       &    $1.810 \pm 0.063$       &    $72.5 \pm 1.0$      \\
         &  &  &   & 20 &   60214.06       &    $1.516 \pm 0.050$       &    $79.3 \pm 0.9$       \\  \hline
3C 120         & 04:33:11.10 &  +05:21:15.62  & 14  & 10 &   60212.15       &    $1.294 \pm 0.109$       &    $100.8 \pm 2.4$      \\
         &  &    &   & 20 &   60214.13       &    $1.387 \pm 0.077$       &    $101.2 \pm 1.6 $      \\\hline
S5 0716+714    & 07:21:53.45 & +71:20:36.36  &  13.5 & 12 &   60211.16     &   $17.015 \pm 0.190$  &   $76.4 \pm 0.3$ \\
    &  &   &  & 10 &   60213.19     &   $9.442 \pm 0.185$  &   $91.0 \pm 0.6$ \\
    &  &   &  & 20 &   60214.17     &   $11.323 \pm 0.132$  &   $97.3 \pm 0.3$ \\\hline
3C 345         & 16:42:58.81 &  +39:48:36.99  &  16.5 & 30 &    60226.86     &   $28.378 \pm 0.542$   &   $47.2 \pm 0.5$   \\ 
    & &        &  & 40 &     60227.86    &   $28.099 \pm 0.331$   &   $42.2 \pm 0.3$   \\  
    & &        &  & 40 &     60228.84    &   $29.938 \pm 0.241$   &   $38.1 \pm 0.2$   \\ 
    & &        &  & 40 &     60229.85    &   $32.092 \pm 0.400$   &   $44.0 \pm 0.4$   \\  \hline
Mrk 501        & 16:53:52.21 &  +39:45:36.61   &  13 & 10 &   60211.81       &    $2.557 \pm 0.082$       &    $121.3 \pm 0.9$      \\
        &  &     &   & 10 &   60212.86       &    $2.798 \pm 0.109$       &    $126.1 \pm 1.1$      \\
        &  &     &   & 10 &   60213.88       &    $2.571 \pm 0.137$       &    $124.1 \pm 1.5$      \\  \hline
OT 081         & 17:51:32.82 &  +09:39:00.73  & 16.5 & 20 &   60211.84       &    $11.574 \pm 0.607$       &    $106.0 \pm 1.5$      \\  \hline
1ES 1959+650 & 19:59:59.85 & +65:08:54.65 &  14.5 & 10 &  60207.95    &  $4.304 \pm 0.144$   &   $159.7 \pm 1.0$   \\   
             & &  &   & 10 &    60210.85       &    $4.515 \pm 0.166$       &    $160.8 \pm 1.1$  \\   
             & &  &   & 10 &    60210.87       &    $4.487 \pm 0.157$       &    $160.6 \pm 1.0$  \\   
             & &  &   & 10 &    60210.89       &    $4.704 \pm 0.124$       &    $160.4 \pm 0.8$  \\   
             & &  &   & 10 &    60210.91       &    $4.576 \pm 0.156$       &    $161.2 \pm 1.0$  \\   
             & &  &   & 10 &    60210.92       &    $4.273 \pm 0.125$       &    $159.9 \pm 0.8$  \\   
             & &  &   & 10 &    60211.88       &    $4.897 \pm 0.131$       &    $159.5 \pm 0.8$   \\ 
             & &  &   & 20 &    60213.92       &    $5.120 \pm 0.105$       &    $160.9 \pm 0.6$  \\    \hline
B 2145+067     & 21:48:05.46 &  +06:57:38.60  &  15.5 & 30 &   60211.93       &    $0.033 \pm 0.288$       &    $142.9 \pm 41.7$      \\  \hline
3C 454.3   & 22:53:57.75 &  +16:08:53.56    & 15.5  & 15 & 60210.97 &    $0.306 \pm 0.878$    &    $66.0 \pm 35.4$      \\  
    & &      &  & 30 & 60211.02 &    $0.479 \pm 0.401$    &    $149.8 \pm 20.0$      \\  \hline
1ES 2344+514   & 23:47:04.84 &  +51:42:17.88   &  14.5 & 20 &   60211.05       &    $1.747 \pm 0.249$       &    $102.1 \pm 4.1$  \\
  &  &     &  & 20 &   60212.01       &    $1.388 \pm 0.189$       &    $104.1 \pm 3.9$ \\  \hline
\enddata
\tablecomments{$^a$Magnitudes reported here refer to typical average magnitudes, indicative of the target brightness/faintness. Blazars are highly variable objects, thus, this value is not representative of the emission of the source at all times. \\
$^{b}$These times refer to the exposure used for each individual image. The total exposure time for a complete 8-cycle observation is 128 times the value reported here.}
\end{deluxetable*}

\begin{figure}
        \centering
        \includegraphics[width=\columnwidth]{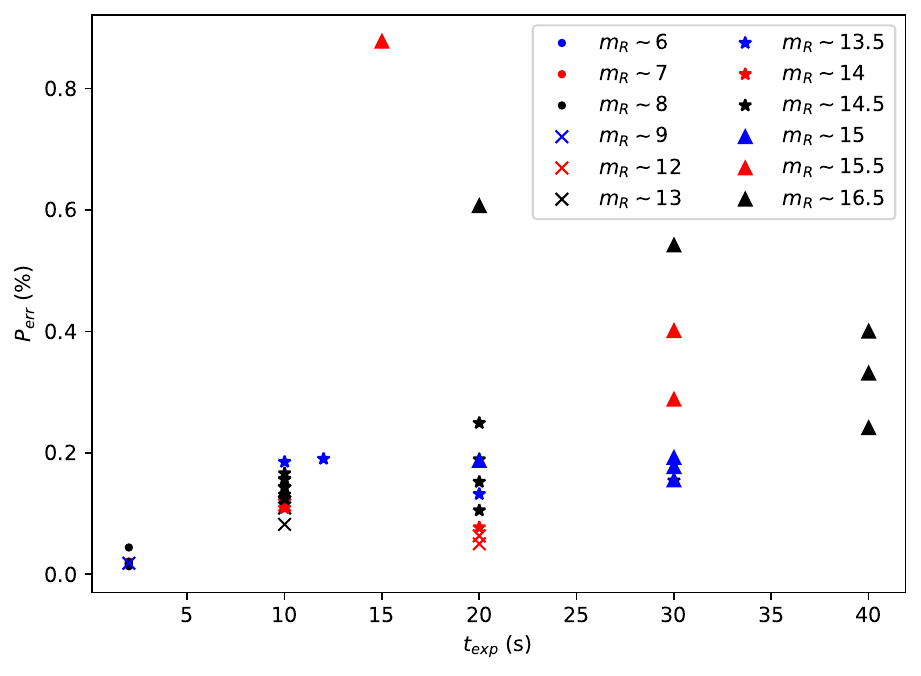}
    \caption{{Polarization degree errors with respect to the exposure time of each image of the 128-image observing cycle. Different markers and colours symbolize different magnitude values.}}
    \label{fig:p_err_vs_t_exp}
\end{figure}

So far the observing time has been mainly dedicated to the observation of blazars under the TOP-MAPCAT monitoring programme for polarimetric observations \citep{agudo2012}. The polarization properties of a sample of blazars observed under this programme are summarized in Table~\ref{tab:blazars_polarization}. Typically, each observation consisted in 8 cycles, with exposures ranging between 10 and 40 seconds, depending on the brightness of the target. This resulted in total observing times of 20 minutes to 1.4 hours. By averaging the Stokes parameters over all the exposure time, precisions $<$0.2\% on the polarization degree and $<$1\degr\ for the polarization angle are achieved for bright blazars (between $m_{R}=12-15.5$). For the faintest targets ($m_{R}=16-16.5$), precisions $<$0.5$-$0.6\% and $<$1.5\degr\ are still reached. {In Figure~\ref{fig:p_err_vs_t_exp} we show the errors obtained for different exposure times and magnitudes, considering the data presented here for high-polarization standards and blazars.} These values represent significant constraints for blazar physics, blazar variability studies and long-term blazar campaigns, proving the very successful performance of DIPOL-1 when observing these targets, especially with the use of a 90 cm telescope. Considering the uncertainties obtained for the faintest blazars of the list and the required total exposure time for reaching such precision, observations of objects fainter than $m_{R}=16.5$ are expected to be relatively time consuming. Nevertheless, and owing to the off-axis autoguider of the T90, these observations are still feasible.

As introduced in Section~\ref{sec4}, the analysis method provides flexibility for dividing the data in shorter temporal intervals, instead of averaging the data from an entire night for each object. Therefore, we have also tested the reliability of measurements on shorter timescales, with the aim of potentially providing time-dense polarimetric observations for future monitoring campaigns. Table~\ref{tab:blazars_polarization_bin} shows the results of the polarization properties for different time bins for the blazars Mrk~501 ($m_{R} \sim 13, t_{exp}=10$~s), J0211+1051 ($m_{R}\sim14.5, t_{exp}=20$~s) and 3C 345 ($m_{R}\sim16.5, t_{exp}=30$~s). In such way, we test the cases of bright, intermediate and faint blazars for which different exposure times were used during their observations. Despite fluctuations introduced by the low number of images used for each estimate, all values are consistent with the one obtained from the night-wise average. Moreover, in all three cases we see a progressive improvement of the uncertainty in polarization degree, as expected. These errors improve up to a factor 4 between the estimates derived from 1 cycle (2.7~min, 5.3~min and 8~min exposures for Mrk~501, 1ES~1950+650 and 3C~345) and 8 cycles (20~min, 40~min and 1~hour exposures for Mrk~501, J0211+1051 and 3C~345), respectively. The same effect is observed for the polarization angle errors, with improvements by a factor $3-4$, depending on the source.

\begin{deluxetable*}{lcccccccccc}
\tablecaption{Polarization properties of the blazars Mrk~501, 1ES~1950+650 and 3C~345 averaging different numbers of 16-image cycles.\label{tab:blazars_polarization_bin}}
\tablehead{
\multirow{3}{*}{Blazar} & \multirow{3}{*}{$m_{R}$$^{a}$} & \multirow{3}{*}{\begin{tabular}[c]{@{}c@{}}Exp. time$^{b}$\\ (s)\end{tabular}} & \multirow{3}{*}{\begin{tabular}[c]{@{}c@{}}N$^{\circ}$ of averaged\\ cycles\end{tabular}} & \multirow{3}{*}{\begin{tabular}[c]{@{}c@{}}Date\\ (MJD)\end{tabular}} &  \multicolumn{2}{c}{$R$} \\  \cline{6-7}    
&  &  & & &  \colhead{$P$ (\%)} & \colhead{$\theta$ (\degr)} 
}
\startdata
Mrk 501 & 13 & 10 & 1 & 60211.81 & $2.933 \pm 0.366$ & $122.3 \pm 3.6$ \\
 & & & 2 & & $2.747 \pm 0.182$ & $120.8 \pm 1.9$ \\
 & & & 3 & & $2.777 \pm 0.131$ & $121.2 \pm 1.4$ \\
 & & & 4 & & $2.732 \pm 0.112$ & $120.9 \pm 1.2$ \\
 & & & 5 & & $2.702 \pm 0.099$ & $120.4 \pm 1.0$ \\
 & & & 6 & & $2.656 \pm 0.089$ & $120.4 \pm 1.0$ \\
 & & & 7 & & $2.604 \pm 0.082$ & $120.5 \pm 0.9$ \\
 & & & 8 & & $2.557 \pm 0.082$ & $121.3 \pm 0.9$ \\
\hline
J0211+1051 & 14.5 & 20 & 1 & 60212.10 & $7.354 \pm 0.481$ & $165.9 \pm 1.9$ \\
 & & & 2 & & $7.243 \pm 0.435$ & $167.8 \pm 1.7$ \\
 & & & 3 & & $7.061 \pm 0.312$ & $166.7 \pm 1.3$ \\
 & & & 4 & & $6.767 \pm 0.243$ & $166.8 \pm 1.0$ \\
 & & & 5 & & $6.701 \pm 0.223$ & $167.0 \pm 1.0$ \\
 & & & 6 & & $6.589 \pm 0.207$ & $166.3 \pm 0.9$ \\
 & & & 7 & & $6.337 \pm 0.173$ & $166.1 \pm 0.8$ \\
 & & & 8 & & $6.301 \pm 0.152$ & $166.1 \pm 0.7$ \\
\hline
3C 345 & 16.5 & 30 & 1 & 60226.86 & $27.815 \pm 1.360$ & $46.6 \pm 1.4$ \\
 & &  & 2 & & $26.993 \pm 1.846$ & $47.1 \pm 0.9$ \\
 & &  & 3 & & $28.173 \pm 0.884$ & $46.9 \pm 0.7$ \\
 & &  & 4 & & $28.561 \pm 0.712$ & $46.8 \pm 0.7$ \\
 & &  & 5 & & $28.955 \pm 0.664$ & $47.1 \pm 0.7$ \\
 & &  & 6 & & $28.976 \pm 0.700$ & $47.4 \pm 0.7$ \\
 & &  & 7 & & $28.410 \pm 0.555$ & $47.1 \pm 0.6$ \\
 & &  & 8 & & $28.378 \pm 0.542$ & $47.2 \pm 0.5$ \\
\enddata
\tablecomments{$^a$Magnitudes reported here refer to typical average magnitudes, indicative of the target brightness/faintness. Blazars are highly variable objects, thus, this value is not representative of the emission of the source at all times. \\
$^{b}$These times refer to the exposure used for each individual image. The total exposure time for a complete 8-cycle observation is 16~$\times$~N$^{\circ}$ of cycles times the value reported here.}
\end{deluxetable*}

Regarding the capability of performing a denser sampling, we observe that for bright blazars ($m_{R}=13-14.5$), we reach fairly good precisions for blazar studies after averaging 2$-$3 cycles, which corresponds to observing times between 5 to 8 minutes for the brightest targets (individual exposures of 10 seconds), and twice the time for 20-second individual exposures. Therefore, DIPOL-1 proves itself to be an incredibly flexible instrument, not only with the ability to perform very high precision polarimetry, but also time-dense intranight polarimetric monitoring on relatively short timescales in the case of bright sources. For faint targets however (e.g. 3C~345, $m_{R}\sim16.5$), the number of cycles needed for obtaining a high precision on the polarization measurements is clearly higher. For such cases it is thus preferred to perform night-wise measurements or, if several hours are invested, establish temporal bins using at least 7 to 8 cycles.

It is worth mentioning that during the still short operation time of DIPOL-1, relevant results are already being obtained. In particular, for the already mentioned blazar 3C~345, we have followed a high $\gamma$-ray emission observed by the \textit{Fermi}-LAT satellite, leading to the detection of an extraordinarily high polarization-degree value of 32\% (see Table~\ref{tab:blazars_polarization}), equaling the all-time highest polarization degree ever detected from this source \citep{smith1986}. Complementing these multi-wavelength enhanced emission states with optical and broadband polarimetry is crucial for understanding the processes leading to the observed variability. In fact, such flaring periods with a fast rise and decay of the optical magnitude and polarization degree have been interpreted under the scenario of shocks travelling through the jet, disturbing the magnetic field. Therefore, observing remarkable events such as the one shown by 3C~345 is key for obtaining final proof of such models.

{Finally, as explained in Section~\ref{sec4}, DIPOL-1 has also the capability of performing photometry when using its complete FoV. We have tested its photometric performance with the TOP-MAPCAT blazar list. We reach typical precisions of $\sim$0.01$-$0.1 magnitudes for $300-600$-s exposures, depending on the source brightness. Therefore, DIPOL-1 is able to complement high-precision polarimetry with photometry. }

\section{Conclusions}\label{sec7}

The new version of DIPOL-1 installed on the T90 has proven its ability of providing fast, time-dense, high-precision polarimetric observations. In this manuscript we have introduced the technical configuration and characteristics of the instrument, as well as the data reduction to be followed for analyzing DIPOL-1 data. Finally, we have demonstrated its remarkable performance with blazar observations, one of the main astrophysical objects to which the T90 is dedicated, through the TOP-MAPCAT monitoring programme \citep{agudo2012}. Here, we summarize the main characteristics and first results obtained with this instrument.

   \begin{enumerate}
      \item We have characterized the instrumental polarization contribution to the observed polarization, with values 4.0806 $\pm$ 0.0014\% and 45.1\degr\ $\pm$ 0.1\degr\ for the instrumental polarization degree and zero-point correction of the position angle, respectively, in the $R$ band. The $G$- and $B$-band values are also reported in Table~\ref{tab:instrumental_polarization} before and after the mirror cleaning. We have ensured the stability of this contribution through the observations of several zero-polarization stars, especially in the $R$ band, owing to its much higher use.
      \item We have been able to accurately measure the polarization of bright, high-polarization standards. The results between different nights with changing atmospheric conditions are in agreement within a 0.05$-$0.1\% polarization degree, explainable through the effect of passing clouds, and  possible small variations of these standards, as reported by \cite{bastien1988}. For such bright targets, we reach precisions on the order of 0.001$-$0.002\%. The stability of the results and calibrations for different zenith distances and pointing directions have also been checked, with successful results. These results provide more updated estimates on the polarization properties of some of the most used polarization stars within the community, that will serve as guide for future polarimetric studies and calibrations.
      \item As first scientific results, we have reported the detection of significant polarized emission from a list of blazars included in the TOP-MAPCAT monitoring programme. By performing a night-wise average of the Stokes parameters, and depending on the brightness of the source (typical magnitudes between $m_{R}=13-16.5$) and the chosen exposure time, we are able to reach uncertainties of $\sim$0.1$-$0.2\% for bright blazars and $\sim$0.5$-$0.6\% for the faintest targets on the list. These are of remarkable precision for blazars observed with a 90 cm telescope.
      \item DIPOL-1 has proven itself to be a reliable instrument for obtaining precise and time-dense polarimetric temporal data series for bright objects. Blazars of magnitudes $m_{R}=13-14.5$ can reach uncertainties of $\sim$0.2$-$0.4\% for the polarization degree and $\sim$1.5\degr\ for the polarization angle with total exposures of 5 to 20 minutes, depending on the brightness and exposure of each individual image.
      \item We have provided optical polarimetric follow-up to a high emission state of the blazar 3C~345 during the first weeks of operation of this instrument, leading in the detection of a consistent high polarization degree ($\sim$28$-$32\%) for this source.
   \end{enumerate}

DIPOL-1 is a joint effort between the University of Turku --- designer, owner and provider of the instrument --- the Instituto de Astrofísica de Andalucía (IAA-CSIC) and the Sierra Nevada Observatory. This paper is based on the commissioning and first observations performed with DIPOL-1 on the T90, thanks to the exceptional labour of the IAA-CSIC and SNO staff.

\begin{acknowledgments}
Based on observations made at the Sierra Nevada Observatory,
operated by the Instituto de Astrof\'isica de Andaluc\'ia (IAA-CSIC).
We thank all SNO staff for their excellent work and support during the installation of the instrument. In particular the support from Francisco Hern\'andez, David P\'erez and Jos\'e Luis de la Rosa was essential.
We also thank the Direction of the IAA-CSIC for the support to the installation, testing and commissioning projects of the DIPOL-1 polarimeter at SNO.  
The IAA-CSIC team acknowledges financial support from the Severo Ochoa grant CEX2021-001131-S funded by the MCIN/AEI/10.13039/501100011033, through grants PID2019-107847RB-C44 and PID2022-139117NB-C44.
DIPOL-1 polarimeters have been built in the cooperation between the University of Turku, Finland, and the Kiepenheuer Institut für Sonnenphysik, Germany, with the support by the Leibniz Association grant SAW-2011-KIS-7, and ERC Advanced Grant Hot-Mol ERC-2011-AdG-291659. P.S-S. acknowledges financial support from the Spanish I+D+i project PID2022-139555NB-I00 funded by MCIN/AEI/10.13039/5011000110033. We thank the referee for her/his thorough review of the manuscript and helpful comments.
\end{acknowledgments}

%

\vspace{5mm}
\facilities{{OSN:0.9m}}


\software{MaxIm DL}




\bibliography{biblio}{}
\bibliographystyle{aasjournal}



\end{document}